\documentclass[preprint,floatfix] {revtex4} 
 
\usepackage{graphicx}
\usepackage{subfigure}
\usepackage{amsmath}
\DeclareMathOperator\arctanh{arctanh}
\begin{document}
\title{Studies on the bound-state spectrum of hyperbolic potential}
\author{Amlan K. Roy}
\altaffiliation{Email: akroy@iiserkol.ac.in, akroy6k@gmail.com. \\                                            
This article is dedicated to my kind-hearted mother Ms.~Radha Rani Roy, on the occasion of her 65th birthday. It is because of her untiring 
effort and constant inspiration that I could pursue research.}
\affiliation{Division of Chemical Sciences, Indian Institute of Science Education and Research (IISER)-Kolkata, 
Mohanpur Campus, P. O. BCKV Campus Main Office, Nadia, 741252, WB, India.}

\begin{abstract}

Bound states of hyperbolic potential is investigated by means of a generalized pseudospectral method. Significantly improved eigenvalues, 
eigenfunctions are 
obtained efficiently for arbitrary $n, \ell$ quantum states by solving the relevant non-relativistic Schr\"odinger equation allowing a non-uniform,
optimal spatial discretization. Eigenvalues accurate up to tenth decimal place are reported for a large range of potential parameters; thus 
covering a wide range of interaction. Excellent agreement with available literature results is observed in all occasions. Special attention is  
paid for higher states. Some new states are given. Energy variations with respect to parameters in the potential are studied in considerable detail 
for the first time. 


\end{abstract}
\maketitle

\section{Introduction}
Ever since the inception of empirical Morse potential about 85 years ago, a vast number of potential functions have been reported
for molecules, with varying degrees of flexibility and accuracy. While much improvements and modifications of the original 
exponential potential is made, the construction of a universal energy-distance relationship for molecules still remains elusive.
Thus there is significant interest and development to mimic the real molecular situations, as evidenced from a large number of 
publications in recent years. Generally, the more the number of parameters in analytic potential energy function, the better it 
fits with experimental data. The literature is vast; some representative recent works include, for example \cite{hajigeorgiou00,
zavitsas03,zavitsas06,coxon06,shayesteh07,coxon10,dattani11} and the references therein.

In this work, we are interested in a three-parameter exponential potential for diatomic molecules, named hyperbolic 
(empirical) potential, suggested in 1986, by Schi\"oberg \cite{schioberg86}, 
\begin{equation}
v(r)= D_e [1-\sigma_0 \coth (\alpha r)]^2, \ \ \ \ \sigma_0 < 1,
\end{equation}
where $D_e, \sigma_0, \alpha$ are three positive adjustable parameters representing the properties of interaction potential.
The potential function has a minimum value zero at a point $r=r_0=\frac{1}{\alpha} \ \arctanh \sigma_0$, where $r_0$ denotes 
equilibrium distance (bond length) between the nuclei. It approaches infinity at the point $r=0$, and goes to
$D_e$ exponentially for large $r$. Also its relation to Morse, Kratzer, Coulomb, harmonic oscillator and other potential functions 
has been discussed \cite{schioberg86}. Further it is suggested that this potential may fit the experimental 
Rydberg-Klein-Rees curve more closely than the Morse function. Some other interesting aspects of this potential can be 
found in the references \cite{schioberg86,lu05}. 

Like many other potentials of physical interest, while the Schr\"odinger equation with hyperbolic potential can be solved for $s$ waves, 
\emph{exact} solutions has not yet been obtained so far for $\ell \ne 0$ states, due to the centrifugal term. Thus approximations are necessary for 
a general eigenstate with arbitrary quantum numbers $n, \ell$. Several attempts have made before, some of which are mentioned now.
Approximate solution of Schr\"odinger equation of diatomic molecules with this potential has been obtained using hypergeometric series 
method \cite{lu05}. Rigorous solutions for $\ell =0$ case have been provided, where the eigenfunctions are expressed in terms of Jacobi 
polynomials \cite{lu05}. Later, arbitrary $\ell$-state solutions are constructed in \cite{dong07} by a proper approximation of the 
centrifugal term offering normalized functions in terms of generalized hypergeometric functions $_2F_1(a,b;c;z)$. Arbitrary $\ell$-wave
solutions have also been suggested by an approximation \cite{dong08} to the centrifugal term as 
$\frac{1}{r^2} \approx \frac{4\alpha^2 e^{-2\alpha r}}{(1-e^{-2\alpha r})^2}$. Good-quality eigenvalues and eigenfunctions for general 
quantum numbers $n, \ell$ were presented by employing a Nikiforov-Uvarov approach \cite{ikhdair09,ikhdair09a,berkdemir09}, such that 
the approximate energy spectra 
and normalized total wave functions are represented in closed form through hypergeometric functions or Jacobi polynomials. In another
development \cite{falaye12}, general eigensolutions are provided by means of an asymptotic iteration method in conjunction with a proper 
approximation to the centrifugal term. Very recently, non-relativistic $\ell$-state solutions of N-dimensional Schr\"odinger equation with
hyperbolic potential has been offered \cite{ortakaya13} in hyperspherical coordinates within the asymptotic iteration method along with 
an approximation to the centrifugal term along the lines of \cite{dong08}. Relativistic bound-state solutions are also discussed by solving
the Dirac equation with an aid of supersymmetric quantum mechanics and functional analysis method \cite{jia07}. Furthermore, it has been 
argued recently \cite{wang12} that the three empirical molecular potentials, \emph{viz.,} Manning-Rosen, Schl\"oberg and Deng-Fan potentials perform 
rather closely to the celebrated Morse potential. This was numerically done in terms of two spectroscopic parameters, namely, vibrational 
rotational coupling parameter $\alpha_e$ and anharmonicity parameter $\omega_e \chi_e$ for a set of 16 selected diatomic molecules. 

The purpose of this work is to make a systematic investigation  on the bound-state spectra of hyperbolic potential through accurate eigenvalues, 
eigenfunction and other properties. It is worth mentioning here that for some of the other molecular potentials like Morse, Kratzer, pseudoharmonic 
or Manning-Rosen, etc., a decent number of theoretical attempts have been published. Thus very good-quality results are available. However for the
potential under consideration, relatively much less effort has been made. In this study, we employ the generalized pseudospectral (GPS) method, 
found to offer promising results for a number of physical situations (see, for example, \cite{roy05,roy05a,sen06,roy07,roy08,roy08a,roy08b,roy13} 
and references therein) in recent years. This is a simple, efficient and yet
accurate method which relies on an optimal, spatial discretization of the radial Schr\"odinger equation. The usefulness and validity of our
proposed scheme is demonstrated through quite accurate ro-vibrational energies of \emph{both low and high-lying} states covering weak, intermediate
and strong interaction in the potential. Thus at first we report
the energies of \emph{both} $s$-wave and rotational states for different $\alpha$ and $\sigma_0$ values. Then a detailed analysis of energy variation
with respect to the parameters $\alpha, \sigma_0, D_e$ is presented, which, to our knowledge, has not been attempted before. A thorough comparison
with literature results is made whenever possible. Section II gives a brief summary of the employed method, while Section IV discusses the 
results. Finally a few concluding remarks are made in Section V. 

\section{The GPS method}
Since this has been already discussed earlier \cite{roy05,roy05a,sen06,roy07,roy08,roy08a,roy08b,roy13}, it suffices here to give only a 
review. The desired radial Schr\"odinger equation in a non-relativistic case can be written as,
\begin{equation}
\hat{H}(r)\ \phi(r) =\varepsilon \ \phi(r).
\end{equation}
The Hamiltonian operator includes usual kinetic and potential energy terms (symbols have their usual meanings),
\begin{equation}
\hat{H}(r) =-\frac{1}{2} \ \ \frac{d^2}{dr^2} +v_{\mathrm{eff}}(r), \ \ \ v_{\mathrm{eff}}(r) = v(r) + \frac{\ell (\ell+1)}{2r^2} 
\end{equation}
and $v(r)$ is given in Eq.~(1). The principal feature of this scheme lies in approximating a function $f(x)$ defined in the interval 
$x \in [-1,1]$ by a polynomial $f_N(x)$ of order N
\begin{equation}
f(x) \cong f_N(x) = \sum_{j=0}^{N} f(x_j)\ g_j(x),
\end{equation}
such that the approximation is \emph {exact} at the \emph {collocation points} $x_j$, i.e., $f_N(x_j) = f(x_j).$
In what follows we employ the Legendre pseudospectral method using $x_0 \! = \! -1$, $x_N \! = \! 1$, where 
$x_j (j=1,\ldots,N-1)$ are obtainable from the roots of first derivatives of Legendre polynomial, $P_N(x)$ 
with respect to $x$, i.e., $P'_N(x_j) = 0.$ $g_j(x)$ in Eq.~(4), called cardinal functions, are given by the following expression,
\begin{equation}
g_j(x) = -\frac{1}{N(N+1)P_N(x_j)}\ \  \frac{(1-x^2)\ P'_N(x)}{x-x_j}.
\end{equation}
They have the unique property $g_j(x_{j'}) \! = \! \delta_{j'j}$. Now the semi-infinite domain $r \in [0, \infty]$ 
is mapped onto a finite domain $x \in [-1,1]$ by a transformation $r=r(x)$. Then one can make use of the 
following algebraic nonlinear mapping, $r=r(x)=L\ \ \frac{1+x}{1-x+\alpha},$ where L and $\alpha=2L/r_{max}$ may be termed as the mapping 
parameters. Furthermore, introducing a symmetrization procedure, eventually leads to the following transformed Hamiltonian, 
\begin{equation}
\hat{H}(x)= -\frac{1}{2} \ \frac{1}{r'(x)}\ \frac{d^2}{dx^2} \ \frac{1}{r'(x)} + v(r(x))+v_m(x), \ \ \ \ 
v_m(x)=\frac {3(r'')^2-2r'''r'}{8(r')^4}.
\end{equation}
The advantage is that this leads to a \emph {symmetric} matrix eigenvalue problem which can be readily 
solved to give accurate eigenvalues and eigenfunctions. Many other features of the method can be found in the references mentioned above.

A series of test calculations were done for a number of potential parameters in the literature to optimize its performance with respect to the 
mapping parameters. In this way, the following parameter set ($r_{max} \! = \! 500, \alpha \! = \! 25, N \! = \! 300$) has been consistently 
used throughout this work, which seemed to be quite satisfactory for the purpose at hand. It has been found that, for lower states, smaller 
values of $r_{max}$ is often sufficient. However for higher states, this needs to be increased presumably to incorporate the complicated 
long-range tail in the wave function. In previous studies with GPS method, such as for Hulth\'en and Yukawa potentials in \cite{roy05a}, similar
phenomenon was observed, where for \emph{higher} excitations and for \emph{stronger} screening parameters, maximum values of $r$ were increased
in order to obtain eigenvalues and eigenfunctions of comparable high accuracy as in the present study. Thus one can obtain decent accuracy with relatively
smaller $r$; however, if the desired accuracy is high, then usually higher values of radial distance is necessary. Recently, a similar situation
has been encountered in the study of bound states of Manning-Rosen potential within a J-matrix approach as well \cite{nasser13}. However, it is 
worthwhile noting here that the accuracy in the GPS method is apparently not affected by the total \emph{number of grid points}, as long as a 
reasonably decent number of collocation points are used for sampling the radial mesh. Thus changing the maximum $r$ has no bearing on the 
computational effort, which is dictated by the dimension of Hamiltonian matrix, i.e., the number of grid points.

\begingroup
\squeezetable
\begin{table}
\caption {\label{tab:table1} Comparison of the calculated eigenvalues (in a.u.) of hyperbolic potential for some selected
states, with $D_e=10$ and $\sigma_0=0.1$. PR signifies Present Result. See text for details.}
\begin{ruledtabular}
\begin{tabular}{ccclccl}
State & $\alpha$ & \multicolumn{2}{c}{Energy} &  $\alpha$ &  \multicolumn{2}{c}{Energy} \\ 
\cline{3-7} 
      &      & PR            & Literature                           &      & PR            & Literature \\   \hline
$1s$  & 0.05 & 1.0454567586  &                                      & 
        0.1  & 1.9750000000$^{\dagger}$  &                                                                          \\
      & 0.15 & 2.7988075454  &                                      & 
        0.2  & 3.5267499380  &                                                                          \\
      & 0.25 & 4.1682605827  &                                      & 
        0.3  & 4.7322309905  &                                                                          \\
$2s$  & 0.05 & 2.6489764589  &                                      & 
        0.1  & 4.3644444444$^{\dagger}$  &                                                                          \\
      & 0.15 & 5.5249158695  &                                      & 
        0.2  & 6.3339868537  &                                                                          \\
      & 0.25 & 6.9089129102  &                                      & 
        0.3  & 7.3209901372  &                                                                          \\
$2p$  & 0.05 & 1.2498051847  &                                      & 
        0.1  & 2.6188810876$^{\dagger}$  & 2.61556\footnotemark[1],2.61874\footnotemark[2],2.61886\footnotemark[3],2.61935\footnotemark[4]  \\ 
      & 0.15 & 3.9057399878  & 3.8983\footnotemark[1],3.90544\footnotemark[2],3.90571\footnotemark[3],3.90645\footnotemark[4]   &
        0.2  & 5.0037812585  & 4.99062\footnotemark[1],5.00331\footnotemark[2],5.00379\footnotemark[3],5.00457\footnotemark[4]  \\ 
      & 0.25 & 5.8865089432  & 5.86611\footnotemark[1],5.88594\footnotemark[2],5.88669\footnotemark[3],5.88725\footnotemark[4]  &
        0.3  & 6.5710530358  &                                                                                                  \\ 
$3s$  & 0.05 & 3.8425598233  &                                      & 
        0.1  & 5.7733673469$^{\dagger}$  &                                                                          \\
      & 0.15 & 6.8428380841  &                                      & 
        0.2  & 7.4628330847  &                                                                          \\
      & 0.25 & 7.8212976304  &                                      & 
        0.3  & 8.0137358269  &                                                                          \\
$3p$  & 0.05 & 2.7996362716  &                                      & 
        0.1  & 4.7355222463$^{\dagger}$  & 4.73223\footnotemark[1],4.73540\footnotemark[2],4.73552\footnotemark[3],4.73638\footnotemark[4]  \\ 
      & 0.15 & 6.0455923499  & 6.03829\footnotemark[1],6.04543\footnotemark[2],6.04570\footnotemark[3],6.04649\footnotemark[4]  &
        0.2  & 6.9165993412  & 6.90394\footnotemark[1],6.91663\footnotemark[2],6.91711\footnotemark[3],6.91733\footnotemark[4]  \\ 
      & 0.25 & 7.4830625708  & 7.46417\footnotemark[1],7.48400\footnotemark[2],7.48475\footnotemark[3],7.48358\footnotemark[4]  &
        0.3  & 7.8362759610  &                                                                                                  \\ 
$3d$  & 0.05 & 1.6309013969  &                                      & 
        0.1  & 3.6274020405  & 3.61747\footnotemark[1],3.62699\footnotemark[2],3.62734\footnotemark[3],3.62769\footnotemark[4]  \\ 
      & 0.15 & 5.2948288570  & 5.27263\footnotemark[1],5.29404\footnotemark[2],5.29485\footnotemark[3],5.2951\footnotemark[4]  &
        0.2  & 6.4757876581  & 6.43684\footnotemark[1],6.47492\footnotemark[2],6.47635\footnotemark[3],6.47598\footnotemark[4]  \\ 
      & 0.25 & 7.2550414560  & 7.19574\footnotemark[1],7.25516\footnotemark[4]                                                  &
        0.3  & 7.7474942818  &                                                                                                  \\ 
$4s$  & 0.05 & 4.7516054891  &                                      & 
        0.1  & 6.6549999999$^{\dagger}$  &                                                                          \\
      & 0.15 & 7.5322784969  &                                      & 
        0.2  & 7.9384186119  &                                                                          \\
$4p$  & 0.05 & 3.9564953219  &                                      & 
        0.1  & 6.0029405462$^{\dagger}$  & 5.99969\footnotemark[1],6.00287\footnotemark[2],6.00299\footnotemark[3],6.0039\footnotemark[4]  \\ 
      & 0.15 & 7.1151441415  & 7.10812\footnotemark[1],7.11526\footnotemark[2],7.11553\footnotemark[3],7.11589\footnotemark[4]   &
        0.2  & 7.7178104204  & 7.70634\footnotemark[1],7.71903\footnotemark[2],7.71951\footnotemark[3],7.71826\footnotemark[4]  \\ 
$4d$  & 0.05 & 3.0818932921  &                                      & 
        0.1  & 5.3316081211  & 5.32177\footnotemark[1],5.33129\footnotemark[2],5.33164\footnotemark[3],5.33216\footnotemark[4]  \\ 
      & 0.15 & 6.7360244502  & 6.71441\footnotemark[1],6.73583\footnotemark[2],6.73663\footnotemark[3],6.73642\footnotemark[4]  &
        0.2  & 7.5431001876  & 7.50672\footnotemark[1],7.54480\footnotemark[2],7.54623\footnotemark[3],7.54331\footnotemark[4]  \\ 
$4f$  & 0.05 & 2.1422543827  &                                      & 
        0.1  & 4.6904242338  & 4.67061\footnotemark[1],4.68965\footnotemark[2],4.69036\footnotemark[3],4.69058\footnotemark[4]  \\ 
      & 0.15 & 6.4310230099  & 6.38708\footnotemark[1],6.42992\footnotemark[2],6.43153\footnotemark[3],6.43112\footnotemark[4]  &
        0.2  & 7.4332936425  & 7.35782\footnotemark[1],7.43397\footnotemark[2],7.43683\footnotemark[3],7.43334\footnotemark[4]  \\ 
$5s$  & 0.1  & 7.2258641975$^{\dagger}$  &                                      & 
        0.2  & 8.0934772157  &                                                                          \\
$5p$  & 0.1  & 6.8034458362$^{\dagger}$  & 6.80027\footnotemark[1],6.80345\footnotemark[2],6.80357\footnotemark[3],6.80432\footnotemark[4]  &
        0.2  & 8.0379547447  & 8.02919\footnotemark[1],8.03813\footnotemark[4]                          \\ 
$5d$  & 0.1  & 6.3777790319  & 6.3681\footnotemark[1],6.37762\footnotemark[2],6.37798\footnotemark[3],6.37842\footnotemark[4]  &
        0.2  & 7.9859391139  & 7.95561\footnotemark[1],7.98606\footnotemark[4]              \\ 
$5f$  & 0.1  & 5.9811548812  & 5.96159\footnotemark[1],5.98063\footnotemark[2],5.98134\footnotemark[3],5.98147\footnotemark[4]  &
        0.2  & 7.9619825982  & 7.89634\footnotemark[1],7.96203\footnotemark[2]                                                  \\ 
$5g$  & 0.1  & 5.6291843445  & 5.59631\footnotemark[1],5.62805\footnotemark[2],5.62924\footnotemark[3],5.62926\footnotemark[4]  &
        0.2  & 7.9681883484  & 7.8515\footnotemark[1],7.9682\footnotemark[2]                                                    \\ 
$6s$  & 0.05 & 6.0116914952  &                                      & 
        0.1  & 7.5999999999$^{\dagger}$  &                                                                          \\
$6p$  & 0.05 & 5.5257326177  &                                      & 
        0.1  & 7.3240501180$^{\dagger}$  & 7.32099\footnotemark[1],7.32416\footnotemark[2],7.32428\footnotemark[3],7.32476\footnotemark[4]  \\
$6d$  & 0.05 & 5.0053978196  &                                      & 
        0.1  & 7.0481354433  & 7.03872\footnotemark[1],7.04824\footnotemark[2],7.04859\footnotemark[3],7.04873\footnotemark[4]  \\
$6f$  & 0.05 & 4.4619148584  &                                      & 
        0.1  & 6.7949021656  & 6.77575\footnotemark[1],6.79479\footnotemark[2],6.79550\footnotemark[3],6.79528\footnotemark[4]  \\
$6g$  & 0.05 & 3.9065692045  &                                      & 
        0.1  & 6.5743566106  & 6.54204\footnotemark[1],6.57377\footnotemark[2],6.57496\footnotemark[3],6.57452\footnotemark[4]  \\
$6h$  & 0.05 & 3.3496137229  &                                      & 
        0.1  & 6.3881065784  &                                                                          \\                   
\end{tabular}
\end{ruledtabular}
\begin{tabbing}
$^{\mathrm{a}}${Ref.~\cite{dong07}.}  \hspace{75pt} \= 
$^{\mathrm{b}}${Ref.~\cite{ikhdair09}.}  \hspace{75pt} \= 
$^{\mathrm{c}}${Ref.~\cite{falaye12}.}  \hspace{75pt} \= 
$^{\mathrm{d}}${Ref.~\cite{lucha99}}, as quoted in \cite{dong07}.  \\
$^{\dagger}$Numerical calculation by the anonymous referee completely reproduces these energies.  \\

\end{tabbing}
\end{table}
\endgroup

\begingroup
\squeezetable
\begin{table}
\caption {\label{tab:table2} Comparison of the calculated eigenvalues (in a.u.) of hyperbolic potential for some selected
states, with $D_e=10$ and $\sigma_0=0.2$. PR signifies Present Result. See text for details.}
\begin{ruledtabular}
\begin{tabular}{ccclccl}
State & $\alpha$ & \multicolumn{2}{c}{Energy} &  $\alpha$ &  \multicolumn{2}{c}{Energy} \\ 
\cline{3-7} 
      &      & PR            & Literature                           &      & PR            & Literature \\   \hline
$1s$  & 0.05 & 0.5206159050  &                                      & 
        0.1  & 1.0099883720  &                                                                          \\
      & 0.15 & 1.4693690763  &                                      & 
        0.2  & 1.9000000000  &                                                                          \\
      & 0.25 & 2.3031087802  &                                      & 
        0.3  & 2.6799042864  &                                                                          \\
$2s$  & 0.05 & 1.4233497848  &                                      & 
        0.1  & 2.5388212990  &                                                                          \\
      & 0.15 & 3.4214960600  &                                      & 
        0.2  & 4.1244444444  &                                                                          \\
      & 0.25 & 4.6860146130  &                                      & 
        0.3  & 5.1344560926  &                                                                          \\
$2p$  & 0.05 & 0.5759729144  &                                      & 
        0.1  & 1.2088815437  & 1.20559\footnotemark[1],1.20876\footnotemark[2],1.20888\footnotemark[3],1.20903\footnotemark[4]  \\ 
      & 0.15 & 1.8666012343  & 1.85922\footnotemark[1],1.86636\footnotemark[2],1.86663\footnotemark[3],1.86689\footnotemark[4]   &
        0.2  & 2.5203736065  & 2.50731\footnotemark[1],2.52000\footnotemark[2],2.52048\footnotemark[3],2.5208\footnotemark[4]  \\ 
      & 0.25 & 3.1471214590  & 3.12683\footnotemark[1],3.14666\footnotemark[2],3.14740\footnotemark[3],3.14766\footnotemark[4]  &
        0.3  & 3.7302842235  &                                                                                                  \\ 
$3s$  & 0.05 & 2.1873069608  &                                      & 
        0.1  & 3.6502178901  &                                                                          \\
      & 0.15 & 4.6502615096  &                                      & 
        0.2  & 5.3383673469  &                                                                          \\
      & 0.25 & 5.8065203909  &                                      & 
        0.3  & 6.1133217893  &                                                                          \\
$3p$  & 0.05 & 1.4701906465  &                                      & 
        0.1  & 2.6831615461  & 2.6799\footnotemark[1],2.68308\footnotemark[2],2.68320\footnotemark[3],2.68358\footnotemark[4]  \\ 
      & 0.15 & 3.6713683033  & 3.66413\footnotemark[1],3.67127\footnotemark[2],3.67154\footnotemark[3],3.67198\footnotemark[4]  &
        0.2  & 4.4650859376  & 4.45247\footnotemark[1],4.46516\footnotemark[2],4.46564\footnotemark[3],4.46579\footnotemark[4]  \\ 
      & 0.25 & 5.0916526920  & 5.07247\footnotemark[1],5.09231\footnotemark[2],5.09305\footnotemark[3],5.09235\footnotemark[4]  &
        0.3  & 5.5749958113  &                                                                                                  \\ 
$3d$  & 0.05 & 0.6846288856  &                                      & 
        0.1  & 1.5790711085  & 1.56921\footnotemark[1],1.57873\footnotemark[2],1.57908\footnotemark[3],1.5792\footnotemark[4]  \\ 
      & 0.15 & 2.5483797803  & 2.52631\footnotemark[1],2.54773\footnotemark[2],2.54853\footnotemark[3],2.54859\footnotemark[4]  &
        0.2  & 3.4820289250  & 3.44311\footnotemark[1],3.48119\footnotemark[2],3.48262\footnotemark[3],3.48228\footnotemark[4]  \\ 
      & 0.25 & 4.3116040028  & 4.25156\footnotemark[1],4.31185\footnotemark[4]                                                  &
        0.3  & 5.0067467188  &                                                                                                  \\ 
$4s$  & 0.05 & 2.8373593986  &                                      & 
        0.1  & 4.4695767827  &                                                                          \\
      & 0.15 & 5.4349868378  &                                      & 
        0.2  & 5.9949999999  &                                                                          \\
$4p$  & 0.05 & 2.2271911909  &                                      & 
        0.1  & 3.7569554439  & 3.75375\footnotemark[1],3.75692\footnotemark[2],3.75704\footnotemark[3],3.75758\footnotemark[4]  \\ 
      & 0.15 & 4.8120058045  & 4.80501\footnotemark[1],4.81215\footnotemark[2],4.81242\footnotemark[3],4.81274\footnotemark[4]   &
        0.2  & 5.5302072359  & 5.51842\footnotemark[1],5.53111\footnotemark[2],5.53159\footnotemark[3],5.53807\footnotemark[4]  \\ 
$4d$  & 0.05 & 1.5621837561  &                                      & 
        0.1  & 2.9528062837  & 2.94305\footnotemark[1],2.95257\footnotemark[2],2.95293\footnotemark[3],2.95317\footnotemark[4]  \\ 
      & 0.15 & 4.1042488047  & 4.08268\footnotemark[1],4.10410\footnotemark[2],4.10491\footnotemark[3],4.1047\footnotemark[4]  &
        0.2  & 5.0009436574  & 4.96371\footnotemark[1],5.00179\footnotemark[2],5.00322\footnotemark[3],5.00137\footnotemark[4]  \\ 
$4f$  & 0.05 & 0.8426504320  &                                      & 
        0.1  & 2.0740692799  & 2.05438\footnotemark[1],2.07342\footnotemark[2],2.07413\footnotemark[3],2.07417\footnotemark[4]  \\ 
      & 0.15 & 3.3572824896  & 3.31338\footnotemark[1],3.35622\footnotemark[2],3.35783\footnotemark[3],3.35742\footnotemark[4]  &
        0.2  & 4.4747398160  & 4.39793\footnotemark[1],4.47408\footnotemark[2],4.47694\footnotemark[3],4.47486\footnotemark[4]  \\ 
$5s$  & 0.1  & 5.0775188943  &                                      & 
        0.2  & 6.3108641975  &                                                                          \\
$5p$  & 0.1  & 4.5494225580  & 4.54628\footnotemark[1],4.54946\footnotemark[2],4.54958\footnotemark[3],4.55015\footnotemark[4]  &
        0.2  & 6.0977741801  & 6.08749\footnotemark[1],6.09822\footnotemark[4]                                                  \\ 
$5d$  & 0.1  & 3.9568482193  & 3.94725\footnotemark[1],3.95677\footnotemark[2],3.95713\footnotemark[3],3.9574\footnotemark[4]  &
        0.2  & 5.8304426454  & 5.79634\footnotemark[1],5.83083\footnotemark[4]                                                  \\ 
$5f$  & 0.1  & 3.3153750988  & 3.29593\footnotemark[1],3.31497\footnotemark[2],3.31568\footnotemark[3],3.31567\footnotemark[4]  &
        0.2  & 5.5607458025  & 5.48854\footnotemark[1],5.56096\footnotemark[4]                                                  \\ 
$5g$  & 0.1  & 2.6411615196  & 2.60844\footnotemark[1],2.64017\footnotemark[2],2.64136\footnotemark[3],2.64124\footnotemark[4]  &
        0.2  & 5.3187638283  & 5.19365\footnotemark[1],5.31882\footnotemark[4]                                                  \\ 
$6s$  & 0.05 & 3.8694697469  &                                      & 
        0.1  & 5.5277892423  &                                                                          \\
$6p$  & 0.05 & 3.4223025374  &                                      & 
        0.1  & 5.1375029556  & 5.13446\footnotemark[1],5.13763\footnotemark[2],5.13775\footnotemark[3],5.13824\footnotemark[4]  \\
$6d$  & 0.05 & 2.9386083805  &                                      & 
        0.1  & 4.6991510082  & 4.68977\footnotemark[1],4.69929\footnotemark[2],4.69965\footnotemark[3],4.69979\footnotemark[4]  \\
$6f$  & 0.05 & 2.4197313136  &                                      & 
        0.1  & 4.2266063298  & 4.20751\footnotemark[1],4.22654\footnotemark[2],4.22726\footnotemark[3],4.22706\footnotemark[4]  \\
$6g$  & 0.05 & 1.8674233592  &                                      & 
        0.1  & 3.7335511103  & 3.70128\footnotemark[1],3.73301\footnotemark[2],3.73421\footnotemark[3],3.73378\footnotemark[4]  \\
$6h$  & 0.05 & 1.2837958703  &                                      & 
        0.1  & 3.2323320970  &                                                                          \\                   
\end{tabular}
\end{ruledtabular}
\begin{tabbing}
$^{\mathrm{a}}${Ref.~\cite{dong07}.}  \hspace{75pt} \= 
$^{\mathrm{b}}${Ref.~\cite{ikhdair09}.}  \hspace{75pt} \= 
$^{\mathrm{c}}${Ref.~\cite{falaye12}.}  \hspace{75pt} \= 
$^{\mathrm{d}}${Ref.~\cite{lucha99}}, as quoted in \cite{dong07}.  
\end{tabbing}
\end{table}
\endgroup

\section{Results and Discussion}
At first, our GPS eigenvalues (in a.u.) of hyperbolic potential having angular quantum number $\ell=0-5$ are presented for some low and high vibrational
levels. Tables 1, 2 report such energies
for $\sigma_0=0.1$, 0.2 respectively for both small (short range) as well as large (long range) $\alpha$ values, for a fixed $D_e=10$ in 
both cases. As mentioned earlier, the analytical expression for eigenvalues, eigenfunctions for $\ell=0$ states are given in 
\cite{lu05,dong07}. Here, we provide the numerical values for some representative states through the current GPS scheme, for sake of completeness 
and assessing approximate 
theoretical methods in future. No results are yet available for $l=5$ states; these are reported here for the first time. First systematic and 
definitive results for arbitrary $(n,\ell)$ states were published in \cite{dong07}. Present calculated ro-vibrational energies seem to agree 
with those of \cite{dong07} in the \emph{short} range (small $\alpha$) more than that in the \emph{long} range (large $\alpha$). Similar 
considerations also hold good as radial as well as angular quantum numbers \emph{increase}. With an improved approximation to the centrifugal 
term as below ($c_0$ is a proper shift found by the expansion procedure),  
\begin{equation}
\frac{1}{r^2} \approx 4 \alpha^2 \left[ c_0 + \frac{e^{-2\alpha r}}{1-e^{-2 \alpha r}} + 
\left( \frac{e^{-2\alpha r}}{1-e^{-2 \alpha r}} \right)^2   \right],
\end{equation}
somehow better eigenvalues, wave functions have been obtained in \cite{ikhdair09}, within the Nikiforov-Uvarov formalism. The current energies
show very good agreement with these, especially in the \emph{long} range as well for \emph{higher} excitations. The recent asymptotic iteration 
results \cite{falaye12} employing a very similar expansion for the centrifugal term, generally produces eigenvalues similar in quality with that 
of \cite{ikhdair09} in majority occasions, but occasionally showing slight improvements in other cases. Wherever possible, numerically 
calculated eigenvalues \cite{lucha99} generated from the MATHEMATICA suite of computer program, as reported in \cite{dong07}, are also 
quoted for easy comparison. These are also in general agreement with other reference energies. We are able to present energies accurate 
up to the tenth place of decimal, for which there appears to be no reference available for direct comparison. 

Now we proceed for changes in energy with respect to potential parameters. Figure 1 depicts such variations for $np \ (n=2-7)$ 
states in the left panel in (a), for $D_e=10, \sigma_0=0.1$ for a considerably broad range of $\alpha$. It is noticed that, for a given 
$\ell$ series, energies steadily increase and then attend a maximum constant value as \emph{alpha} reaches a particular value. The initial 
increase becomes progressively sharper for higher $n$. For higher $n$, apparently this maximum in energy is 
attained for smaller $\alpha$ and vice versa. For other $\ell$ series (with varying $n$), similar trend is observed and not produced here 
to save space. In (b) now, the energy change with $\alpha$ is 
shown for 2p state, at seven values of $\sigma_0$, \emph{viz.}, 0.1, 0.2, 0.3, 0.4, 0.5, 0.6 and 0.8 respectively, keeping $D_e$ 
fixed at 10 again. The threshold $\alpha$ value, after which energy becomes constant, is gradually shifted to higher values as $\sigma_0$
is increased. A similar trend has been observed for other states as well, which are not repeated. 

\begin{figure}
\begin{minipage}[c]{0.40\textwidth}
\centering
\includegraphics[scale=0.45]{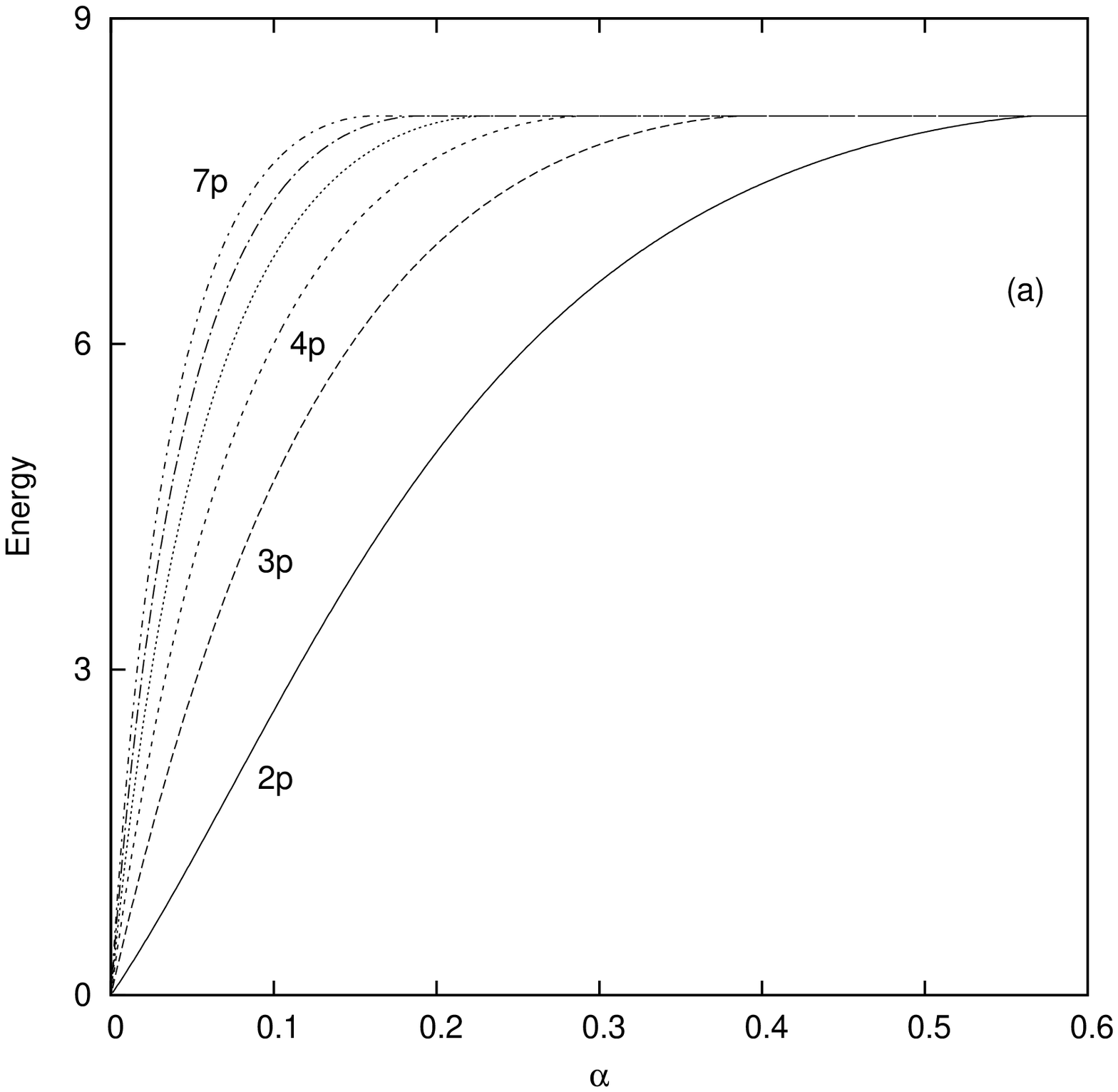}
\end{minipage}%
\hspace{0.5in}
\begin{minipage}[c]{0.40\textwidth}
\centering
\includegraphics[scale=0.45]{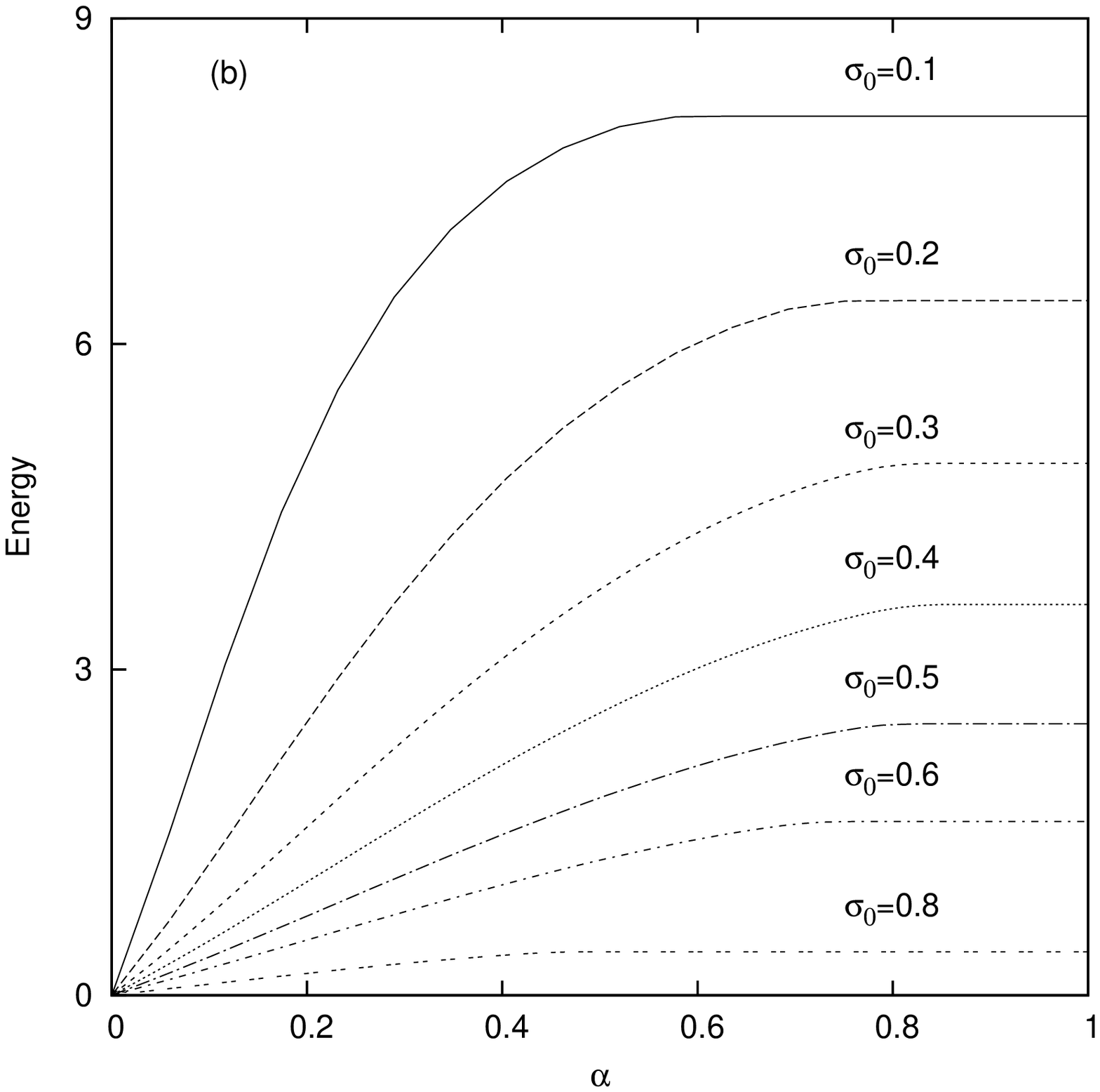}
\end{minipage}%
\caption{Energy eigenvalues (a.u.) of hyperbolic potential as function of $\alpha$. The left panel (a) corresponds to 
$np \ (n=2-7)$ states, having $D_e=10, \sigma_0=0.1$, while (b) shows the same for 2p state for a fixed $D_e=10$, and 
$\sigma_0=0.1, 0.2, 0,3, 0.4, 0.5, 0.6$, 0.8 respectively.}
\end{figure}

In Fig.~2, energy variations are considered with respect to $\sigma_0$, for a constant value of $D_e=10$. The left panel (a) displays this for
$np \ (n=2-7)$ states, having $\alpha=0.1$. For all of them, at first energy sharply falls from an initial fixed value and then starts to decrease 
at a slower rate after some particular $\sigma_0$, resembling much like an exponential decay. For a given $\ell$ quantum number, the initial 
fall-off is much stronger for lower $n$, and its extent gradually diminishes for higher values of $n$, so much so that for 10p (not shown in the 
figure), it very much looks like a linear decay. For other $\ell$ series of the hyperbolic potential, very similar trend is followed and thus 
omitted.  The right panel (b) now shows how energy changes with $\sigma_0$, for a state, for seven values of $\alpha=0.1$, 0.2, $\cdots$, 0.7, 
taking 2p as a representative. For progressively higher values of $\alpha$, the separation between two successive curves tends to diminish and 
individual plots appear to be much flatter, once again approaching a linear behavior for sufficiently higher value of $\alpha$. The general 
qualitative features remain same for other states as well.

\begin{figure}
\begin{minipage}[c]{0.40\textwidth}
\centering
\includegraphics[scale=0.45]{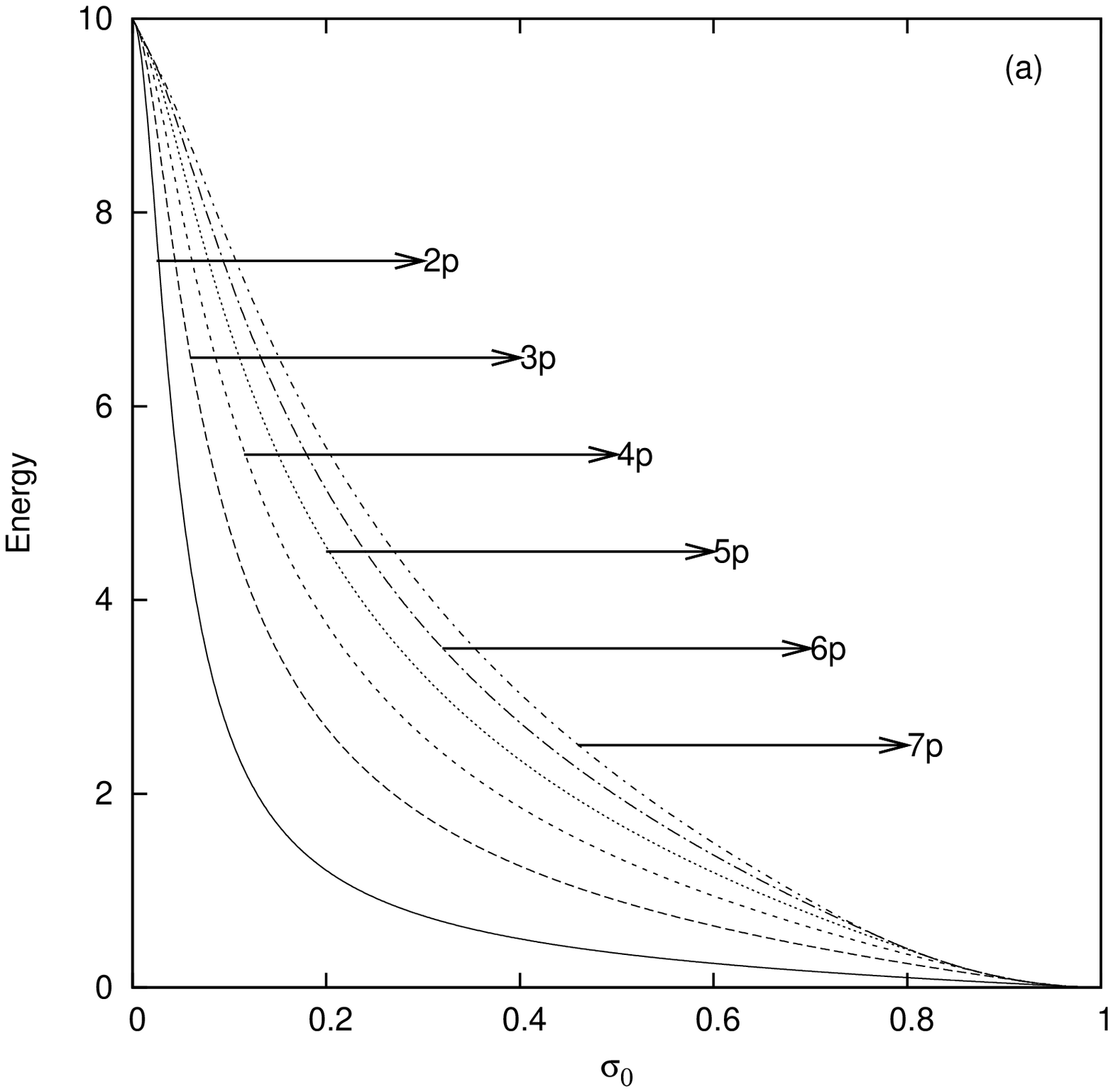}
\end{minipage}%
\hspace{0.5in}
\begin{minipage}[c]{0.40\textwidth}
\centering
\includegraphics[scale=0.45]{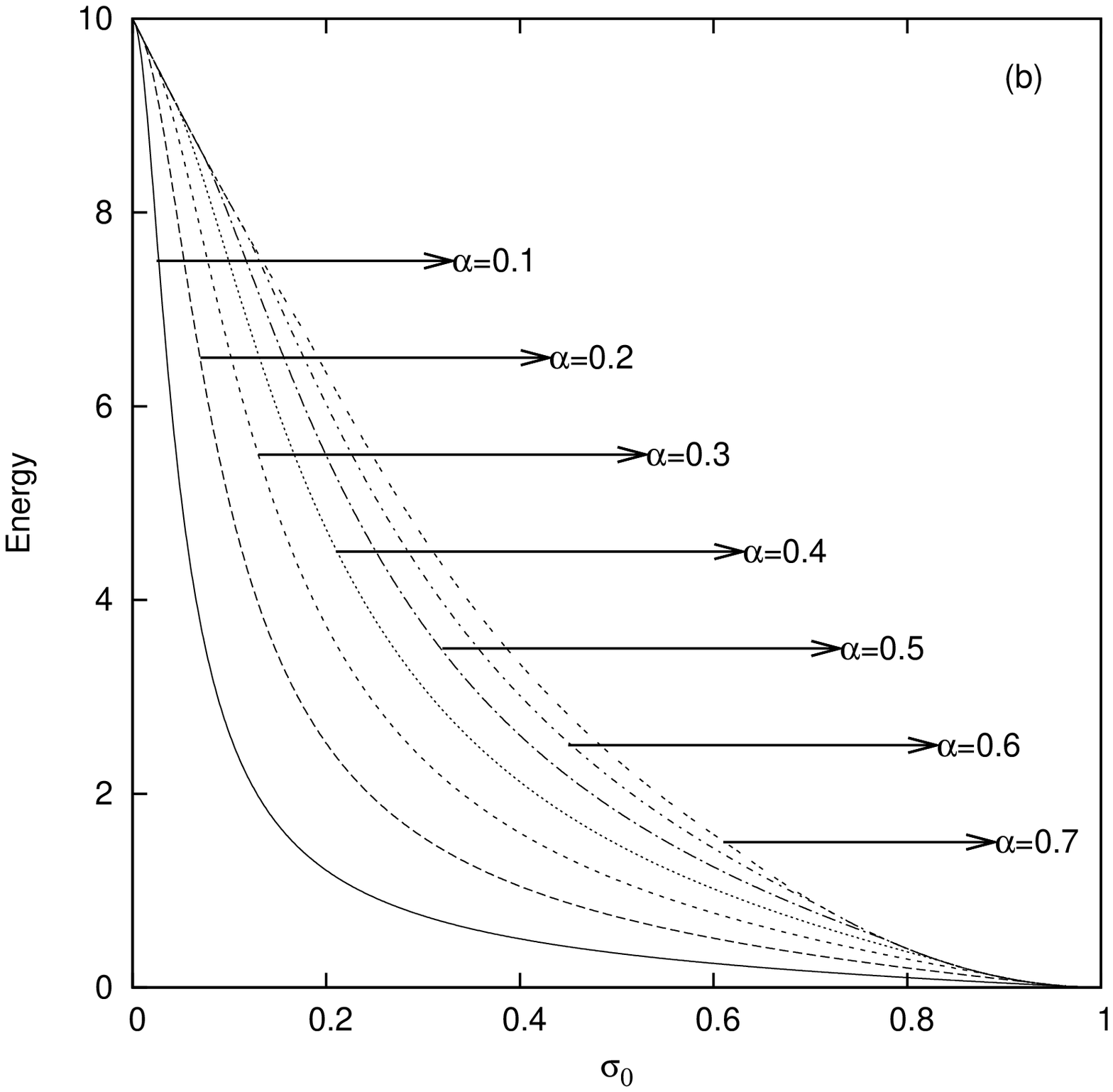}
\end{minipage}%
\caption{Energy eigenvalues (a.u.) of hyperbolic potential as function of $\sigma_0$ for (a) $np \ (n=2-7)$ states, having 
$D_e=10, \alpha=0.1$, (b) 2p state for fixed $D_e=10$, $\alpha=0.1, 0.2, \cdots, 0.7$.}
\end{figure}

Finally Fig.~3 shows energy eigenvalues (in a.u.) with respect to the parameter $D_e$. First in the left panel (a), 2p--8p state energies are 
depicted for a sufficiently large range of $D_e$, keeping both $\alpha, \sigma_0$ constant at 0.1. For low-$n$ states, energy increases
steadily in the beginning at smaller $D_e$, which is eventually arrested after some time, with a consequent slow down of further increase. As $n$ 
assumes higher values, the rate of increase tends to remain unhindered resulting very much a straight-line like shape. Now the middle panel (b) 
shows similar 
plots for a representative 2p state for seven $\alpha$ values of 0.1, 0.2, $\cdots$, 0.7 (covering short and long range), keeping $\sigma_0=0.1$, 
in all cases. Same observation as in (a) holds true; as $\alpha$ goes to higher values, the individual plots lead to higher energy 
values and also tend to become more like straight line. Also the separation between two neighboring $\alpha$ plots becomes smaller and smaller as 
the latter increases. Then in the right panel (c), we give the E versus $D_e$ plots, again for a selected state (2p) for seven $\sigma_0$ values, 
namely, 0.1, 0,2, $\cdots$, 0.7, having a constant $\alpha=0.1$. Note that the energy axis in (c) is different from (a), (b). In this case, for higher 
$\sigma_0$, energies go to lower values, and much like in (a), (b), the separation between two adjacent $\sigma_0$ plots is reduced. 
Very recently, energy variation with $D_e$ for $1s, 2s, 2p, 3s$ states of this potential has been studied (for fixed $\sigma_0=0.2$) in 
\cite{ortakaya13} by means of asymptotic iteration method, which seems to corroborate our present findings. 

\begin{figure}
\begin{minipage}[c]{0.30\textwidth}
\centering
\includegraphics[scale=0.30]{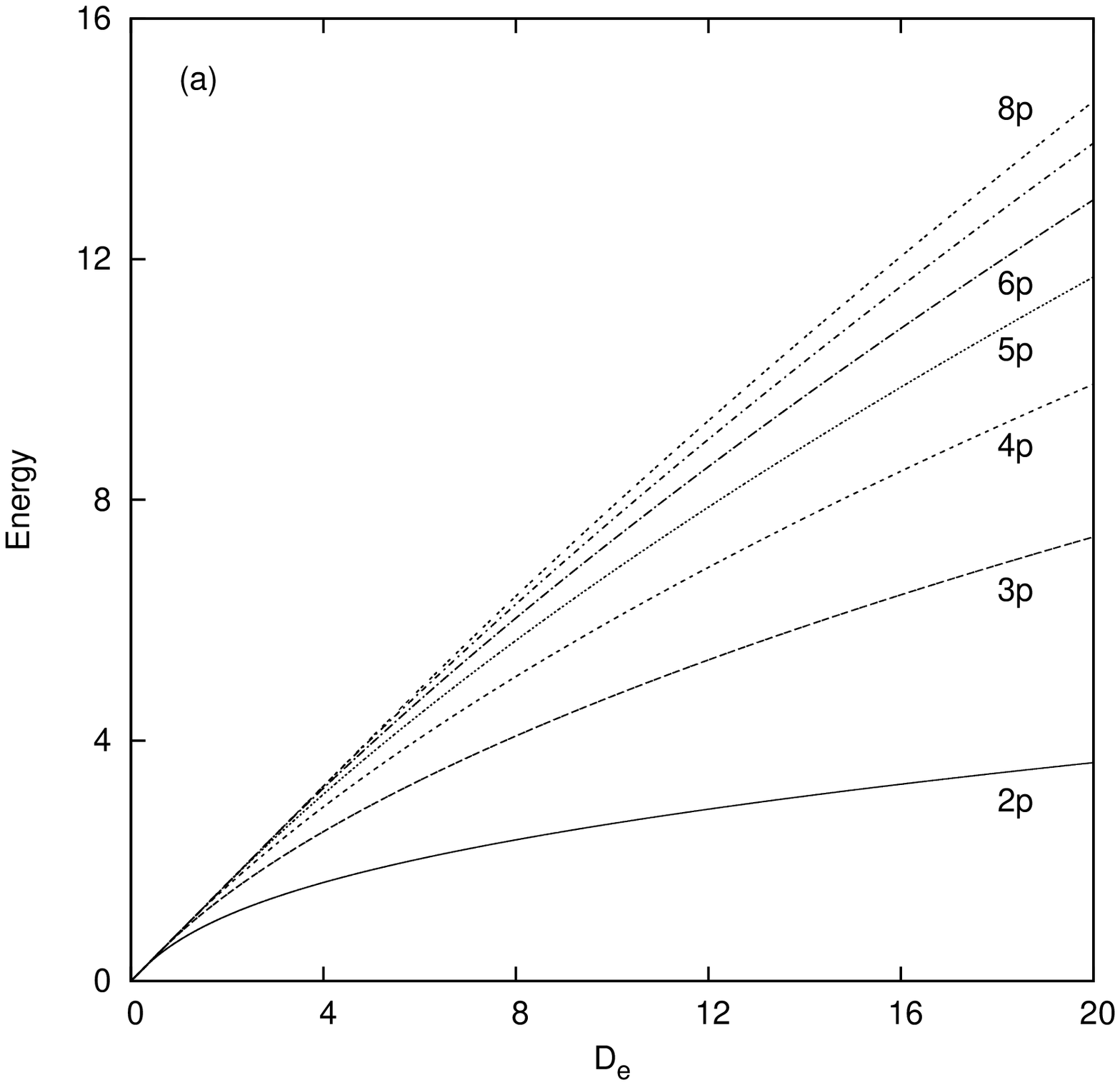}
\end{minipage}%
\hspace{0.1in}
\begin{minipage}[c]{0.30\textwidth}
\centering
\includegraphics[scale=0.30]{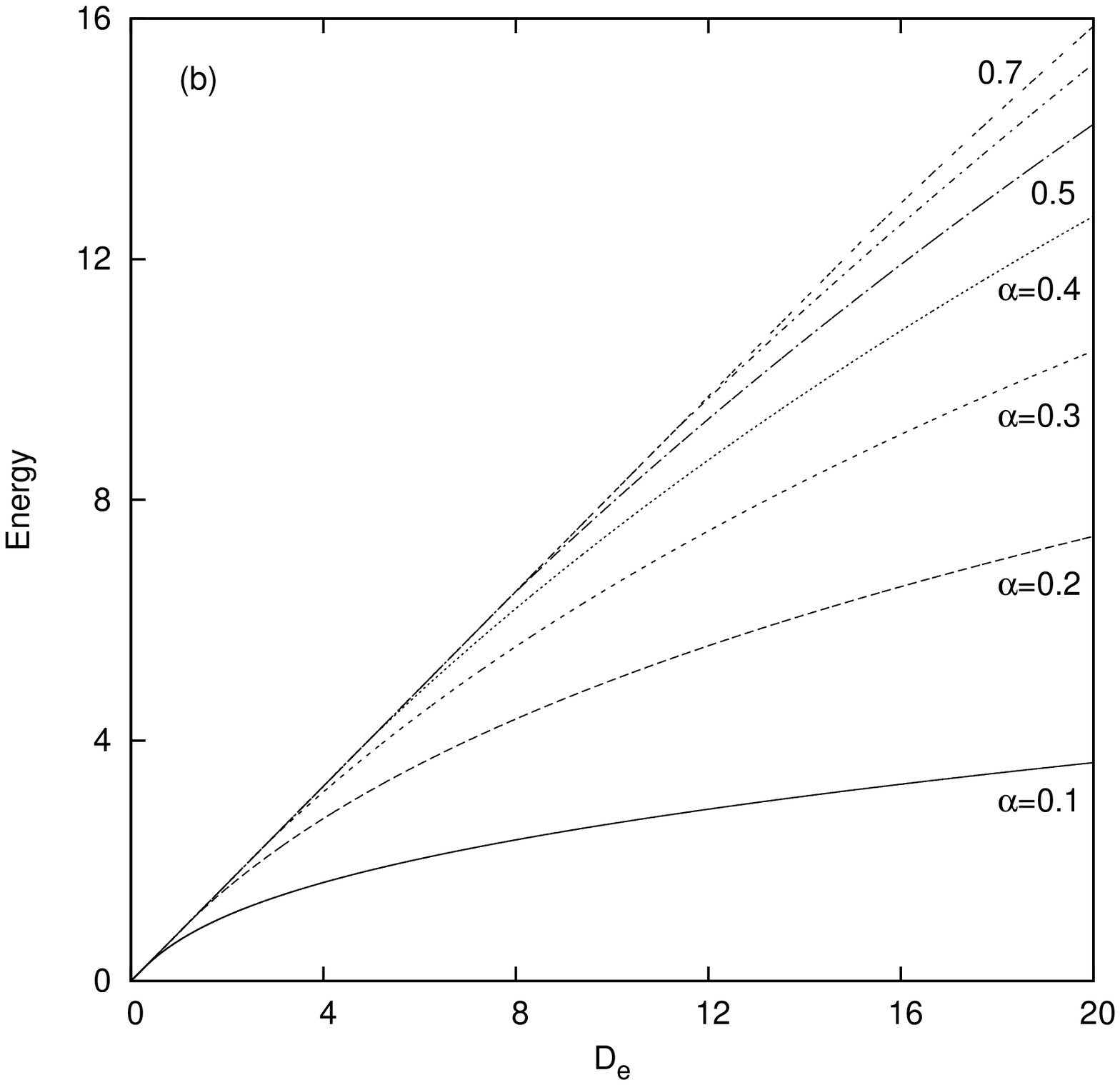}        
\end{minipage}%
\hspace{0.1in}
\begin{minipage}[c]{0.30\textwidth}
\centering
\includegraphics[scale=0.30]{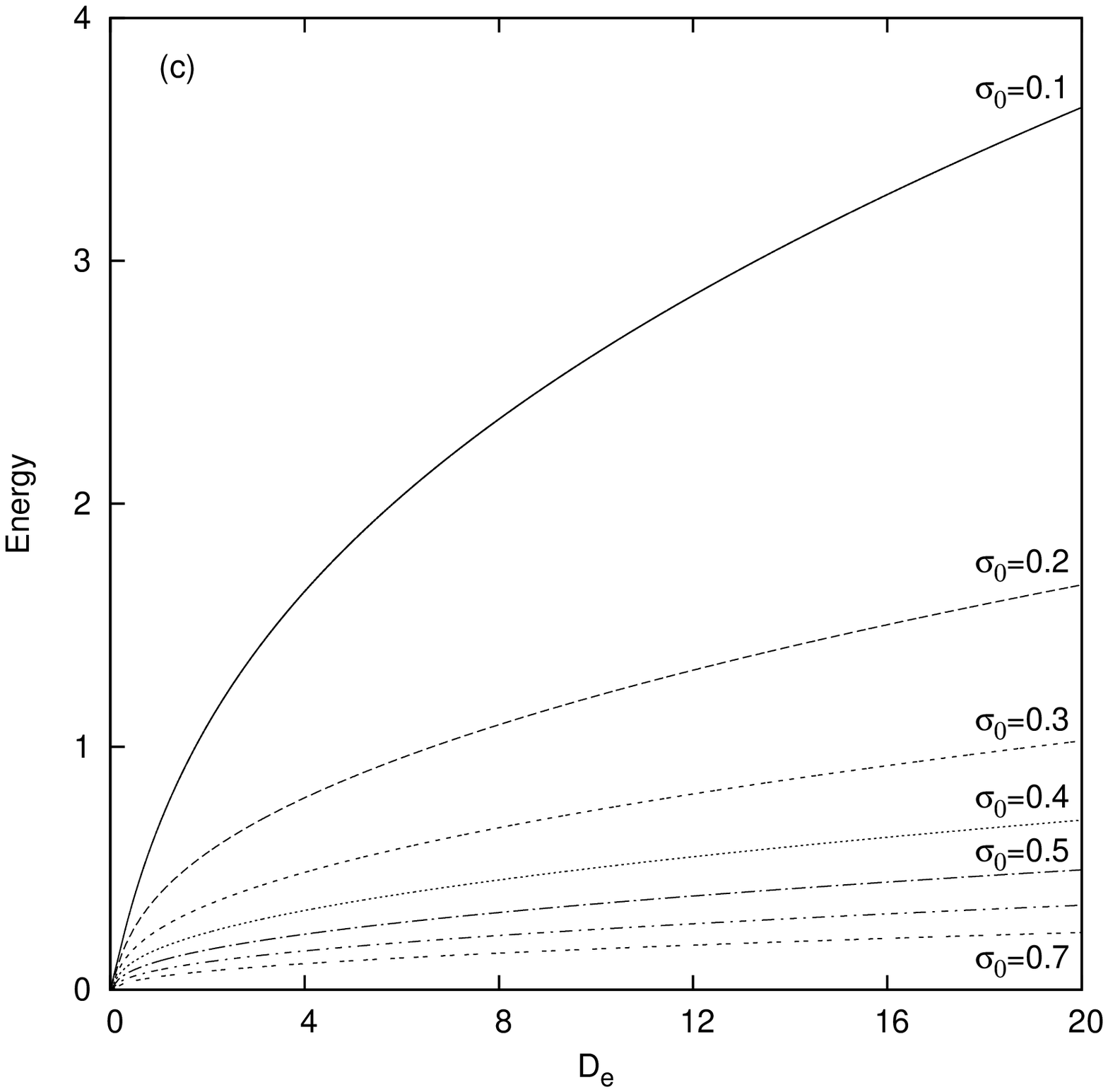}        
\end{minipage}%
\caption{Energy eigenvalues (a.u.) of hyperbolic potential as function of $D_e$, for (a) $np \ (n=2-8)$ states, having 
$\sigma_0= \alpha=0.1$, (b) 2p state with $\sigma_0=0.1$, $\alpha=0.1, 0.2, \cdots, 0.7$,  and (c) 2p state with $\alpha=0.1$ and 
$\sigma_0=0.1, 0.2, \cdots, 0.7$ respectively.}
\end{figure}

\section{conclusion}
In the present communication, we have provided accurate solutions of the Schr\"odinger equation with a hyperbolic potential. Non-relativistic
eigenvalues are obtained easily by means of a generalized pseudospectral method, giving an optimal radial discretization. The validity and 
feasibility of this is demonstrated by producing very good-quality ro-vibrational energies for arbitrary $n,\ell$ quantum numbers. Energies accurate
up to ten to eleven significant figures are reported here for the first time, which compares quite excellently with the existing literature data.
Special emphasis has been given to higher states as well in the long range of the potential.
A detailed investigation on the variation of energy with respect to potential parameters has been made for the first time. Many new states are
reported. In essence, an accurate and reliable scheme for solution of this and similar potential is presented.

\section{acknowledgment} It is a pleasure to thank Mr.~Shahid Ali Farooqui of the Computer Center for his help on making the plots. I am 
grateful to the Director, Prof.~R.~N.~Mukherjee, for his constant support and encouragement. Sincerely thank goes to the anonymous referee for his 
kind, constructive and valuable comments, from which the manuscript has greatly benefited.

\end{document}